 \documentclass[aps,prb,twocolumn,showpacs]{revtex4}

\usepackage{amsmath}
\usepackage{hyperref}
\usepackage{graphicx}
\usepackage{color}
\usepackage{graphics}
\usepackage{epsfig}

\usepackage{color}
\usepackage{amsfonts,amsmath,mathrsfs}
\usepackage{slashed}
\usepackage{epsfig}
\usepackage{latexsym}
\usepackage{bbm}{\large }
\usepackage{amssymb}
\usepackage{graphicx}% Include figure files
\usepackage{dcolumn}% Align table columns on decimal point
\usepackage{bm}% bold math
\begin{document}

%\title{2D Lattice Model Construction of Symmetry Protected Topological Phases}

\title{Projective construction of two-dimensional symmetry-protected topological phases with U(1), SO(3), or SU(2) symmetries}

\author{Peng Ye$^1$ and Xiao-Gang Wen$^{1,2,3}$}
\affiliation{$^1$Perimeter Institute for Theoretical Physics, Waterloo, Ontario, Canada N2L 2Y5\\
$^2$Department of Physics, Massachusetts Institute of Technology, Cambridge, Massachusetts 02139\\
$^3$Institute for Advanced Study, Tsinghua University, Beijing, 100084, P. R. China}
\begin{abstract}
We propose a general approach to construct symmetry protected topological
(SPT) states (\ie the short-range entangled states with symmetry) in
2D spin/boson systems on lattice.  In our approach, we
fractionalize spins/bosons into different fermions, which occupy nontrivial
Chern bands.  After the Gutzwiller projection of the free fermion state
obtained by filling the Chern bands, we can obtain SPT states on
lattice. In particular, we constructed a U(1) SPT state of a spin-1 model, a SO(3) SPT state of a boson system with spin-1 bosons and spinless bosons, and
a SU(2) SPT state of a spin-1/2 boson system. By applying the ``spin gauge field'' which directly couples to the spin density and spin current of $S^z$ components, we also calculate the quantum spin Hall conductance in each SPT state. The projective ground states can be further studied numerically in the future by variational Monte Carlo etc.

\end{abstract}
\date{{\small \today}}
\maketitle
 \section{Introduction}

Recently, quantum entanglement becomes an important concept and a powerful tool
in the fields of quantum information and quantum many-body
systems.\cite{LW0605,KP0604}  A long-range entangled state (LRE) can not be
deformed to a direct product state by any local unitary (LU)
transformation.\cite{Chen10,Chenlong,Wenscience} The fractional quantum Hall
effect (FQHE)\cite{Tsui82} is a concrete example of LRE and has been considered
as an example of intrinsic ``topological order''.\cite{Wtop,WNtop,Wrig}  If a
gapped quantum state is short-range entangled (SRE), one can always deform it
to a spatially direct product state by a LU transformation\cite{Chenlong}.  So
a SRE state is trivial. But if a SRE state is invariant under some symmetry
group $G$, then it can be nontrivial (even if it does not break any symmetry).
This because for some symmetric SRE states, there is no way to locally deform
the states into direct product states unless one explicitly breaks the symmetry
in LU procedure.  We call such SRE states ``symmetry-protected topological
phases'' (SPT).\cite{Chen10,Chenlong,Wenscience}  Topological
insulators\cite{TI} (TI) belongs to fermionic SPT protected by $\mathbb{Z}_2$
time reversal symmetry and U(1) particle number conservation.  On the other
hand,  if a gapped symmetric quantum state is long-range entanglements, it
will be called ``symmetry-enriched topological state'' (SET).

Group theory has been successfully utilized in analyzing symmetry-breaking
phases.  Recently, it was shown that group cohomology with U(1) coefficient can
be utilized in  describing SPT phases systematically.\cite{Chenlong} A concrete
SPT phase is the 1D Haldane phase with SO(3) symmetry where integer spin is a
faithful representation of SO(3) symmetry.    It is known that the spin-1
antiferromagnetic chain is semiclassically (in large spin-$S$ expansion)
described by a nonlinear sigma model (NLSM) with 2$\pi$ topological theta term.
The two boundaries are occupied by two dangling free 1/2 spins which are
protected by SO(3) symmetry and contributes to fourfold degeneracy. 
 
It is interesting that, by using group cohomology theory, one can construct
exactly solvable lattice models which realize specific SPT phases respecting
given symmetry groups\cite{Chen11a,Chenlong,Liu11}. Along this line, some
interesting lattice models of bosonic SPT phases with discrete symmetry groups
have been proposed in literatures. For continuous symmetry group, however, the
solution of cocycle construction in group cohomology is more difficult
mathematically.  2D SPT phases with U(1)
symmetry\cite{LV1219,LS0903} and with SO(3) and with SU(2)
symmetries\cite{Liu12} are studied using Chern-Simons field theory and
Wess-Zumino-Witten (WZW) field theory.\cite{WZ71,Witten83,Witten84} But, a
concrete lattice model for realizing those SPT phases is much desired, which is
more physical than continuum field theory description. 

Instead of directly constructing exactly solvable models, in this paper, we
present an effective approach to realize SPT phases on 2D lattice.  Our
approach is based on the projective construction of strongly correlated bosonic
or spin systems,\cite{BZA8773,BA8880,AM8874,KL8842,SHF8868,%
AZH8845,DFM8826,WWZcsp,Wsrvb,LN9221,MF9400,WLsu2}. In this construction, bosons/spins are
fractionalized into several fermions which form some mean-field states before projection. The projection can be done by considering the confined phase of internal gauge fields which glue the fermions back into bosons/ spins. Recently, bosonic topological insulators (a bosonic SPT with U(1)$\rtimes$Z$^T_2$ symmetry group where U(1) is particle number conservation and Z$^T_2$ is time-reversal symmetry) on 3D lattice are constructed in this way\cite{YW13}. In the present paper, we shall consider 2D SPT systems with continuous symmetry groups. We assume that fermions occupy several independent
nontrivial  Chern bands respectively.  After the Gutzwiller projection of the
free fermion state obtained by filling the Chern bands, we can obtain SPT
states on lattice. We have constructed a U(1) SPT
state, a SO(3) SPT state, and a SU(2) SPT state using the projective
construction.

\section{$U(1)$ SPT states in spin-1 lattice model}
\label{spinF}  

\subsection{mean-field ansatz}

 In this section, we are going to construct a SPT phase with $U(1)$ symmetry on
lattice.\cite{Chenlong,LS0903,LV1219,CW1217} Such a state has been obtained by
several other constructions.\cite{SL1204,GV1207,Lu12b,LL1263}  Our lattice
model is a spin-1 model on a honeycomb lattice, with 3 states, $|m\>$,
$m=0,\pm1$, on each site.  We can view our spin-1 model as spin-1/2
hardcore-boson model with $|m=0\>$ state as the no-boson state (or two bosons with different spins) and $|m=\pm 1\>$
state as the one-boson state with spin-up or spin-down.

In a fermionic projective construction, we write the bosonic operators as
\begin{align}
 b_{\al,i} = f_{\al,i} c^{(s)}_i,\ \ \ \al=\up,\down,\label{fractionalization11}
\end{align}
where, $f_{\uparrow},f_{\downarrow}$ are two fermions with spin-$\uparrow$ and spin-$\downarrow$.  $s=1,2$ is the orbital index of $c$\, fermions. $c_i^{(1)}$ and $c_i^{(2)}$ are annihilation operators of $c$\, fermions in two orbitals of $c$\, fermions at site-$i$, respectively. The nature of hard-core is implied by $(b_\alpha)^2=0$. In this projective construction where bosons do not carry $s$-index, we note that creating a boson with spin-$\alpha$ is identical to creating one f fermion with spin-$\alpha$ and creating one $c$ fermion of any possible orbital. The purpose of enlarging orbital number of $c$ fermions is to satisfy the particle number constraint as we shall show below.

The identity (\ref{fractionalization11}) is not an operator identity. Rather, the matrix elements of the two sides in the physical Hilbert space are identical to each other. The physical Hilbert space we consider is three-dimensional spin-1 space ($|m_i=0,\pm1\>$) at any site-$i$. In terms of fermions, each spin state can be constructed by two different fermionic states:
\begin{align}
&m=0  \text{\,\,\,state\,\,\,}:\,\,    |0\> \text{\,\,\,and\,\,\,} f^{\dagger}_{i\up}c^{(1)\dagger}_i f^{\dagger}_{i\down}c^{(2)\dagger}_i|0\> \,\,,\\ 
 &m=1  \text{\,\,\,state\,\,\,}:\,\,f^\dag_{i\up} c_i^{(1)\dag}|0\>\text{\,\,\,and\,\,\,} f^\dag_{i\up} c_i^{(2)\dag}|0\>\,,\\
& m=-1  \text{\,\,\,state\,\,\,}:\,\,f^\dag_{i\down} c_i^{(1)\dag}|0\>\text{\,\,\,and\,\,\,} f^\dag_{i\down} c_i^{(2)\dag}|0\>\,,
\end{align}
where, $|0\>$ denotes vacuum with no particles. For example, the state at site-$i$ is $|m_i=1\>$ if one $f_\uparrow$ fermion and one $c^{(1)}$-fermion occupy the site, or, if one $f_{\uparrow}$-fermion and one $c^{(2)}$-fermion occupy the site. Therefore, the three-dimensional spin space is actually a six-dimensional Hilbert space in terms of fermions and each of spin states is constructed by equal number of $f$\, fermions and $c$\, fermions, which is the particle number constraint implemented by the present fermionic projective construction. This can be also seen clearly by noting that  the projective construction in Eq. (\ref{fractionalization11}) introduces an internal gauge field $\mathcal{A}_\mu$ that is responsible for gluing $f$\,  and $c$\, fermions together to form a hard-core boson. By choice of convention, both of $f_{\uparrow}$ and $f_\downarrow$ ($c^{(1)}$ and $c^{(2)}$) carry $+1$ ($-1$) gauge charge of $\mathcal{A}_\mu$. As a result, by integrating out the temporal component $\mathcal{A}_0$, the above particle number constraint on particle numbers is obtained.

Therefore, we can start with a many-fermion state of $f_{i,\al}$ and $c_i$, $|\Psi\>$, and obtain a physical spin state $|\Phi\>$ by projecting into the
sub-space with equal $f$\, fermions and $c$\, fermions on each site:
\begin{align}
|\Phi\>= P |\Psi\>. 
\end{align}
Using such a projective construction, we would like to
construct a U(1) SPT state on a honeycomb lattice. The projection operator $P$ is given by:
\begin{align}
P=&\prod_i \bigg(|m_i=+1\>\big(\<i,\up\,,(1)|+\<i,\uparrow\,,(2)|\big)\nonumber\\
&+|m_i=-1\>\big(\<i,\down\,,(1)|+\<i,\down\,,(2)|\big)\nonumber\\
&+|m_i=0\>\big(\<i,0|+\<i,\uparrow,(1),\downarrow,(2)|\big)\bigg)
\end{align}
where, $|i,\alpha\,, (s)\>$ denotes $f^\dagger_{i\alpha}c^{(s)\dagger}_i|0\>$.

Let us consider the following free fermion Hamiltonian on a honeycomb lattice
\begin{align}
 H_{MF}&=\sum_{i,j} 
          \Big[
\mathbf{f}_{j}^\dag U_{ji} \mathbf{f}_{i}
+\mathbf{c}_{j}^\dag V_{ji} \mathbf{c}_{i}
\Big]\,,
\end{align}
where, $\mathbf{f}\equiv (f_{\uparrow},f_{\downarrow})^T$,  $\mathbf{c}\equiv (c^{(1)},c^{(2)})^T$, and, $U_{ji}$ are $2\times 2$ matrices satisfying
\begin{align}
 U^\dag_{ji} = U_{ij},
\end{align}
and $V_{ji}$:
\begin{align}
 V^\dagger_{ji} = V_{ij}\,.
\end{align}
Let $|\Psi\>$ be the lowest energy state of the above free fermion Hamiltonian.
Then $|\Phi\>=P|\Psi\>$ will be a spin/boson state induced by the above free
fermion Hamiltonian.
%\begin{align}
%\Phi\left(\left\{m_i \right\}\right)=\langle 0|\prod_{i}\mathcal{O}_{i}|\Psi\>\,,
%\end{align}
%where, $m_i=0,\pm1$, $\left\{m_i\right\}$ defines a complete Ising basis of Hilbert space of original spin-1 model. The operator $\mathcal{O}_i$ is  defined as follows. If $m_i=0$, $\mathcal{O}_i=\mathbb{I}$ where $\mathbb{I}$ is identity operator; if $m_i=1$, $\mathcal{O}_i=f_{i\uparrow} c^{(1)}_i$ or $=f_{i\uparrow} c^{(2)}_i$; if $m_i=-1$, $\mathcal{O}_i=f_{i\downarrow} c^{(1)}_i+f_{i\downarrow} c^{(2)}_i$.  It is easy to check that before projection, the Hilbert space dimension of each site is $2^{4}=16$. After projection, the physical spin Hilbert space at each site is reduced to 3, which is spanned by $|0\>\,,f^\dagger_{\uparrow}(c^{(1)\dagger}+c^{(2)\dagger})|0\>\,, f^\dagger_{\downarrow}(c^{(1)\dagger}+c^{(2)\dagger})|0\>$.
We say that such a spin/boson state is described by the
ansatz $U_{ji}$ and $V_{ji}$.  
%We can show that if the ansatz has the form
%\begin{align}
% U^\dag_{ji}= 
%\imth u^0_{ji}\si^0
%+ u^l_{ji}\si^l, \ \ \ \ u^0_{ji},u^l_{ji}=\text{ real}
%\end{align}
%then both $|\Psi\>$ and $|\Phi\>$ will be spin singlet.
%Here $\si^0$ is the $2\times 2$ identity matrix and
%$\si^l$, $l=1,2,3$, are  the Pauli matrices.

To construct a SPT state using the above projective construction,
we choose the ansatz $U_{ij}$ to be\cite{KM0501}
\begin{align}
 U_{ij}&= t_0\si^0 , \ \ \ ij = \text{ nearest neighbour (NN) links}
\\
 U_{ij}&= \imth \nu_{ij} t \si^3 \,,\, 
ij 
=\text{next nearest neighbour (NNN) links},
\nonumber 
\end{align}
where, the complex number $v_{ij}$ is
\begin{align}
 v_{ij}&=t_0 , \ \ \ ij = \text{NN links}
\\
 v_{ij}&= \imth \nu_{ij} t , \ \ \ 
ij 
= \text{NNN links},
\nonumber 
\end{align}
Here, $t_0,t$ are hopping energies of NN and NNN respectively. $\sigma^0$ is a two-dimensional identical matrix, $\sigma^{1,2,3}$ are three usual Pauli matrices. 
$\nu_{ij}=+$ if the fermion $f$ makes a right turn going from $j$ to $i$ on the
honeycomb lattice, and $\nu_{ij}=-$ if the fermion $f$ makes a left turn. In this ansatz, each of $f_{\up}\,,f_{\down}$ contains two bands (due to the two independent sites per unit cell of the honeycomb lattice) with a gap at
zero energy.  Each lower band has a Chern number $+1$ for the $f_{\up}$-fermions
and a Chern number $-1$ for the $f_{\down}$-fermions.  Let's fill the lower band of $f_\uparrow$ and the lower band of $f_\uparrow$ completely, such that each unit cell is filled with one $f_\uparrow$ and one $f_\downarrow$ fermion respectively. For c-fermions, each unit cell contains four bands due to two independent sites of honeycomb lattice and two orbitals at each site. In order that the particle number constraint between f and c fermions is satisfied at least in mean-field level, an appropriate design of the two by two matrix $V_{ij}$ leads that the two lower bands are filled and the total Chern number of these two bands (i.e. summation of the two Chern numbers of the two bands) is equal to $+1$. Now, each unit cell contains one $f_\uparrow$, one $f_\downarrow$, and two $c$ fermions.
  
%For later convenience, let's correlate fermion's operators to physical spin operators of underlying spin-1 model on the honeycomb lattice. The physical spin-1 operators ($S^z,S^+,S^-$) of original spin model can be expressed in terms of fermions by recovering all matrix elements of the spin operators in the spin-1 space: $S^z=n_c(n_\uparrow-n_\downarrow)$ ($n_\uparrow, n_\downarrow, n_c$ are fermionic particle number operators), $S^+\equiv\frac{1}{\sqrt{2}}(S^x+iS^y)=c^\dagger f^\dagger_{\uparrow}+c f_{\downarrow}$, $S^-\equiv\frac{1}{\sqrt{2}}(S^x-iS^y)=c^\dagger f^\dagger_{\downarrow}+cf_{\uparrow}$. Since all these operators work in the three-dimensional projected subspace, $S^z$ can be replaced by $S^z=P( n_\uparrow-n_\downarrow) P$, which does not alter its matrix elements. We can also check that in the projected subspace, $(S^+)^2=P f^\dagger_{\uparrow}f_{\downarrow}P$, $(S^-)^2=Pf^\dagger_{\downarrow}f_{\uparrow}P$, $(S^z)^2=P(n_\uparrow+n_\downarrow )P$. Therefore, $P n_{\uparrow}P =((S^z)^2+S^z)/{2}$, $P n_\downarrow P=((S^z)^2-S^z)/{2}$, $P n_cP =(S^z)^2$, $P f^\dagger_\uparrow cP=P f^\dagger_\downarrow cP=0$. 
 
\subsection{Effective field theory}
In the
following, we will show that $|\Phi\>=P|\Psi\>$ is a bosonic SPT state
protected by the $U(1)$ symmetry (generated by $S^z$).
First, from the  free fermion Hamiltonian, we see that the ground state state
$|\Psi\>$ respects the $S_z$ spin rotation symmetry generated by $S^z$.
The $f_\al$-fermions form a ``spin Hall'' state described by
$U(1)\times U(1)$ Chern-Simons theory
\begin{align}
 \cL&=\frac{1}{4\pi}(
a_{\up\mu}\prt_\nu a_{\up\la}\eps^{\mu\nu\la}
-
a_{\down\mu}\prt_\nu a_{\down\la}\eps^{\mu\nu\la}
)
\nonumber\\
&\ \ \ \ \ \ \
+\frac{1}{2\pi}
A^\text{spin}_{\mu}\prt_\nu ( a_{\up\la} -a_{\down\la}) \eps^{\mu\nu\la}
\nonumber\\
&=
\frac{1}{4\pi} K_{IJ} a_{I\mu}\prt_\nu a_{J\la}\eps^{\mu\nu\la}
+\frac{1}{2\pi} q_{I} A^\text{spin}_{\mu}\prt_\nu a_{I\la}\eps^{\mu\nu\la}
\end{align}
with $I,J=1,2$ and
\begin{align}
 K=\bpm 1 &0 \\
        0 &-1 \epm, \ \ \ \ 
 q=\bpm 1  \\
        -1 \epm.
\end{align}
The coupling gauge charges of $f_\up$ and $f_\down$ in $q$-vector are $+1$ and $-1$ since $_\up$ and $_\down$ are related to the two spinful states of the physical spin-1 operator at each site.    
Here the ``spin gauge field'' $A^\text{spin}_\mu$ is the gauge potential that {\it directly} couples to
the $S_z$ spin density and current, which is different from the definition of conventional spin Hall effect where the applied external gauge field is the electromagnetic field. Therefore, this ``spin Hall'' mean-field ansatz for $f$ fermions has no spin Hall conductance ( $\frac{1}{2\pi}q^T K^{-1}q=0$) in the present definition, although it is nonzero ($\frac{1}{2\pi}(1,1)K^{-1}q=2\times\frac{1}{2\pi}$) if external electric field (composed by electromagnetic gauge potential) is the probe field instead. Since the U(1) global symmetry which defines the U(1) SPT discussed here is the spin rotational symmetry, it is physically natural to gauge this U(1) global symmetry (the corresponding gauge potential is nothing but $A_\mu^{\text{spin}}$) and discuss its response theory to diagnose SPT physics.  In the whole paper, we shall apply the spin gauge field $A_\mu^{\text{spin}}$ to probe the {\it intrinsic} quantum spin Hall response phenomenon.

The $c$\, fermions form an ``integer quantum Hall'' state described by
$U(1)$ Chern-Simons theory
\begin{align}
 \cL&=
\frac{1}{4\pi} b_{\mu}\prt_\nu b_{\la}\eps^{\mu\nu\la}
\end{align}
Thus the total effective theory is given by
\begin{align}
\cL&=
\frac{1}{4\pi} K_{IJ} a_{I\mu}\prt_\nu a_{J\la}\eps^{\mu\nu\la}
+\frac{1}{2\pi} q_{I} A^\text{spin}_{\mu}\prt_\nu a_{I\la}\eps^{\mu\nu\la}
\nonumber\\
&\ \ \ \ \ \
+\frac{1}{4\pi} b_{\mu}\prt_\nu b_{\la}\eps^{\mu\nu\la}
\end{align}
The projection $P$ is done by setting the total $f_\al$-fermion
density-current, $J^f_\mu= \frac{1}{2\pi} \prt_\nu (a_{\up\la} +a_{\down\la})
\eps^{\mu\nu\la} $, equal to the $c$\, fermion density-current, $J^c_\mu=
\frac{1}{2\pi} \prt_\nu b_{\la}  \eps^{\mu\nu\la} $.\cite{Wtoprev,Wpcon}  After setting
$b_{\la}=(a_{\up\la} +a_{\down\la})$, we reduce the effective theory  to
\begin{align}
\label{U1SPT}
\cL&=
\frac{1}{4\pi} \bar K_{IJ} a_{I\mu}\prt_\nu a_{J\la}\eps^{\mu\nu\la}
+\frac{1}{2\pi} q_{I} A^\text{spin}_{\mu}\prt_\nu a_{I\la}\eps^{\mu\nu\la}
\end{align}
with
\begin{align}
\bar K=\bpm 2 &1 \\
       1 &0 \epm, \ \ \ \
\bar K^{-1}=\bpm 0 &1 \\
       1 &-2 \epm. 
%q=\bpm 1  \\
%       -1  \epm .
\end{align}
Since $|\text{det}(\bar K)|=|-1|=1, $\Eqn{U1SPT} is the low energy effective
theory for the spin/boson state $|\Phi\>=P|\Psi\>$ after the projection.  Such
a  low energy effective theory describes a $U(1)$ SPT state with an integer
Hall conductance (U(1) is spin rotation symmetry of spin-1 along z-axis)
\begin{align}
 \si_{xy}=q^T\bar K^{-1}q\frac{1}{2\pi}
 =-4\frac{1}{2\pi}.
\end{align}
We note that the spin Hall conductance is an even integer
in the unit of $\frac{1}{2\pi}$.

\section{SO(3) SPT states in a lattice model with spin-1 bosons and spinless bosons}\label{fractionalization}
\subsection{mean-field ansatz}
SO(3) SPT phases were firstly studied in Ref.\onlinecite{Liu12} where field theoretic approach is applied. The bulk is described by SO(3) principal chiral nonlinear sigma model (the field variable $g$ is $3\times3$ SO(3) matrix) with 2$\pi K$ topological theta term where $K$ (${K}/{4}\in\mathbb{Z}$) completely labels distinct SPT phases. The boundary is described by a SO(3)$_K$ WZW term where SO(3) transformation is defined as left-multiplication, \ie $g\rightarrow hg$ with $h\in$SO(3)$_L$, such that only left mover carries SO(3) charge and right mover is SO(3) charge-neutral. Due to this chiral SO(3) transformation, the boundary can not be simply replaced by a SO(3)$_K$ WZW critical spin chain that is nonchiral. 

In this section, we are going to construct a SPT phase protected by $SO(3)$
symmetry on lattice.  Our lattice model contain spin-1 bosons and three spin-0
bosons on a honeycomb lattice. Therefore, this model is a model of many-boson, such as a cold-atom system in 2D.

 In a fermionic projective construction, we
write the spin-1 bosonic operators (labeled by a spin index $m=0,\pm1$) as
\begin{align}
 b_{m,i} = f_{m,i} c^{(s)}_i,\ \ \ m=0,\pm 1\label{so31}
\end{align}
and the three ($m=0,\pm 1$) spin-0 bosons as
\begin{align}
 \t b_{m,i} = \t f_{m,i} c^{(s)}_i,\ \ \ m=0,\pm 1\label{so32}
\end{align}
Here, $s=1,\cdots, 6$ is  the orbital index for $c$\, fermions at each site.  
 
Again, we can start with a many-fermion state of $f_{m,i}$, $\t f_{m,i}$ and $c_i$,
$|\Psi\>$, and obtain a physical spin state $\Phi\>$ by projecting into the
sub-space where the sum
of the $f$\, fermion number 
and the $\t f$-fermion number is equal to 
the $c$\, fermion number on each site:
\begin{align}
|\Phi\>= P |\Psi\>. 
\end{align}
Using such a projective construction, we can
construct a SO(3) SPT state on a honeycomb lattice.

Let us consider the following free fermion Hamiltonian on a honeycomb lattice
\begin{align}
 H&=\sum_{i,j} 
          \Big[
\mathbf{f}_{j}^\dag u_{ji} \mathbf{f}_{i} 
+\widetilde{\mathbf{f}}_{j}^\dag u^*_{ji} \widetilde{\mathbf{f}}_{i}, 
+\mathbf{c}_{j}^\dag V_{ji} \mathbf{c}_{i}, 
\Big]
\end{align}
where, $\mathbf{f}=(f_{+1},f_{0},f_{-1})^T\,,\widetilde{\mathbf{f}}=(\widetilde{f}_{+1},\widetilde{f}_{0},\widetilde{f}_{-1})^T\,,\mathbf{c}=(c^{(1)},c^{(2)},c^{(3)},c^{(4)},c^{(5)},c^{(6)})^T$, and, 
$u_{ji}$ 
are complex numbers satisfying
\begin{align}
 u^*_{ji} = u_{ij}, \ \ \ \ 
\end{align}

To construct a SO(3) SPT state using the above projective construction,
we choose the ansatz to be
\begin{align}
 u_{ij}&=1 , \ \ \ ij = \text{NN links}
\\
 u_{ij}&= \imth \nu_{ij} t , \ \ \ 
ij 
= \text{NNN links},
\nonumber 
\end{align}
where
$\nu_{ij}=+$ if the fermion $f$ makes a right turn going from $j$ to $i$ on the
honeycomb lattice, and $\nu_{ij}=-$ if the fermion $f$ makes a left turn.  
For such an ansatz, each of  the six fermions ($f_{m}$ and $\widetilde{f}_{m}$) forms two energy bands with a gap at zero energy and the lower one is filled. Each lower band  has a Chern number $+1$ for the $f$\, fermions and a Chern number $-1$ for the $\t f$-fermions.  

The 6 by 6 matrix $V_{ij}$ leads to a 12 by 12 single particle Hamiltonian in momentum space for $c$ fermions in the honeycomb unit cell. Let's fill the six bands from the bottom and assume that the band gap exists between the sixth and seventh bands. The total Chern numbers contributed by the six bands is equal to $+1$. As a result, in each unit cell, there are three $f$ fermions (one for each of the three components), three $\widetilde{f}$ fermions (one for each of the three components), and six $c$ fermions, which satisfy the particle number constraint.

\subsection{Effective field theory}

 In the
following, we like to show that $|\Phi\>=P|\Psi\>$ will be a bosonic SPT state
protected by the SO(3) symmetry (generated by $f^\dag T^l f$, where $T^l$,
$l=1,2,3$, are the $3\times 3$ matrices that generate the SO(3) Lie algebra). From the  free fermion Hamiltonian, we see that the ground state state
$|\Psi\>$ respects the SO(3) spin rotation symmetry generated by $f^\dag T^l
f$. 
 The $f$\, fermions form an ``integer quantum Hall'' state described by
$U^3(1)$ Chern-Simons theory
\begin{align}
 \cL&=
\frac{1}{4\pi} K_{IJ} a_{I\mu}\prt_\nu a_{J\la}\eps^{\mu\nu\la}
+\frac{1}{2\pi} q_{I}A^{\text{spin}}_{\mu}\prt_\nu a_{I\la}\eps^{\mu\nu\la}
\end{align}
with $I,J=1,2,3$ and
\begin{align}
 K=\bpm 1 &0 & 0\\
        0 &1 & 0\\
        0 &0 & 1 \epm, \ \ \ \ 
 q=\bpm 1  \\
        0 \\
        -1 \epm
.
\end{align}
Here $A^{\text{spin}}_\mu$ is the gauge potential that couple to the $S_z$ spin density
and current.  The $\t f$-fermions also form an ``integer quantum Hall'' state
with an opposite Hall conductance, which is described by $U^3(1)$ Chern-Simons
theory
\begin{align}
 \cL&=-
\frac{1}{4\pi}  K_{IJ} \t a_{I\mu}\prt_\nu \t a_{J\la}\eps^{\mu\nu\la}
\end{align}
The $c$\, fermions form an ``integer quantum Hall'' state described by
$U(1)$ Chern-Simons theory
\begin{align}
 \cL&=
\frac{1}{4\pi} b_{\mu}\prt_\nu b_{\la}\eps^{\mu\nu\la}
\end{align}
Thus the total effective theory is given by
\begin{align}
\cL&=
\frac{1}{4\pi} K_{IJ} a_{I\mu}\prt_\nu a_{J\la}\eps^{\mu\nu\la}
-\frac{1}{4\pi} K_{IJ} \t a_{I\mu}\prt_\nu \t a_{J\la}\eps^{\mu\nu\la}
\nonumber\\
&\ \ \ \ 
+\frac{1}{2\pi} q_{I} A^{\text{spin}}_{\mu}\prt_\nu a_{I\la}\eps^{\mu\nu\la}
+\frac{1}{4\pi} b_{\mu}\prt_\nu b_{\la}\eps^{\mu\nu\la}
\end{align}
The projection $P$ is done by setting the total $f$\, fermion
and
 $\t f$-fermion
density-current, $J^f_\mu= \sum_I \frac{1}{2\pi} \prt_\nu (a_{I\la} +\t a_{I\la})
\eps^{\mu\nu\la} $, equal to the $c$\, fermion density-current, $J^c_\mu=
\frac{1}{2\pi} \prt_\nu b_{\la}  \eps^{\mu\nu\la} $.  After setting
$b_{\la}=\sum_I (a_{I\la} +\t a_{I\la})$, we reduce the effective theory  to
\begin{align}
\label{SO3SPT}
\cL&=
\frac{1}{4\pi} \bar K_{IJ} \bar a_{I\mu}\prt_\nu \bar a_{J\la}\eps^{\mu\nu\la}
+\frac{1}{2\pi} \bar q_{I} A^{\text{spin}}_{\mu}\prt_\nu \bar a_{I\la}\eps^{\mu\nu\la}
\end{align}
with $I,J=1,2,3,4,5,6$ and
\begin{align}
\bar K=\bpm 
2 & 1 & 1 & 1 & 1 & 1 \\
1 & 2 & 1 & 1 & 1 & 1 \\
1 & 1 & 2 & 1 & 1 & 1 \\
1 & 1 & 1 & 0 & 1 & 1 \\
1 & 1 & 1 & 1 & 0 & 1 \\
1 & 1 & 1 & 1 & 1 & 0 \\
       \epm, \ \ \ \
\bar q=\bpm 
1\\
0\\
-1\\
0\\
0\\
0
       \epm.
\end{align}
Also $\bar a_{I,\mu}=a_{I\mu}$ and $\bar a_{I+3,\mu}=\t a_{I\mu}$,
$I=1,2,3$.  Since $|\text{det}(K)|=|-1|=1$, 
\Eqn{SO3SPT} is the low energy effective theory for the spin-1 boson state
$|\Phi\>=P|\Psi\>$ after the projection.
Such a  low energy effective theory describes a
SO(3) SPT state with an integer Hall conductance
for the $S_z$ ``charge''
\begin{align}
 \si_{xy}=\bar q^T\bar K^{-1}\bar q\frac{1}{2\pi}
 =2\frac{1}{2\pi}.
\end{align}
Again, we note that the spin Hall conductance is an even integer
in the unit of $\frac{1}{2\pi}$. This nontrivial quantum spin Hall effect indicates that gapless edge Hall current is protected by U(1) subgroup of SO(3) spin rotational symmetry. Beyond this K matrix formulation, Ref. (\onlinecite{Liu12}) proved that edge can not be further gapped out by  keeping non-abelian SO(3) / SU(2) symmetry, because the mass term which gaps out the excitations will mix the left mover and right mover and hence breaks the SO(3) / SU(2) symmetry. To realize this result in our projective construction, we need to further construct the same WZW theory description for edge states. We shall discuss SU(2) case in Sec. \ref{sectionKM} and SO(3) is similar.

\section{SU(2) SPT state in a spin-1/2 boson model on lattice}
\subsection{mean-field ansatz}
SU(2) SPT phases were also firstly studied in Ref.\onlinecite{Liu12} where field theoretic approach is applied. The bulk is described by SU(2) principal chiral nonlinear sigma model (the field variable $g$ is $2\times2$ SU(2) matrix) with 2$\pi K$ topological theta term where integer $K$ labels different SPT phases. The boundary is described by a SU(2)$_K$ WZW term where SU(2) transformation is defined as left-multiplication, \ie $g\rightarrow hg$ with $h\in$SU(2)$_L$, such that only left mover carries SU(2) charge and right mover is SU(2) charge-neutral. Due to this chiral SU(2) transformation, the boundary can not be simply replaced by a SU(2)$_K$ WZW critical spin chain that is nonchiral. 

In the following, we are going to construct a SPT phase with a SU(2)
symmetry on lattice. Our lattice model contains spin-1/2 
bosons.  Using the fermionic projective construction, we
write the spin-1/2 bosonic operators (labeled by a spin index $\alpha$ and a flavor index $s$) as
\begin{align}
 b_{\al,i} = f_{\al,i} c^{(s)}_i\,,\ \ \ \al=\up,\down
\end{align}
where, $s$ is the orbital index of $c$\, fermion at each site. Let's consider a general construction. $N_O$ ($N_{O}=2$ is fixed in the following discussion) and $N_L$ denote the orbital number per site  and independent site number per unit cell, respectively. At each site, $c$ fermions can be viewed as a two-dimensional ``pseudo-spinor''. For honeycomb lattice, $N_L=2$. 
Again, we can start with a many-fermion state of $f_{\al,i}$
and $c_i$,
$|\Psi\>$, and obtain a physical spin state $|\Phi\>$ by projecting into the
sub-space where the 
the $f$\, fermion number
is equal to 
the $c$\, fermion number on each site:
\begin{align}
|\Phi\>= P |\Psi\>. 
\end{align}
Using such a projective construction, we can
construct a SU(2) SPT state on lattice.

Let us consider the following free fermion Hamiltonian
\begin{align}
 H&=\sum_{i,j} 
          \Big[
\mathbf{f}_{\al j}^\dag u_{ji} \mathbf{f}_{\al i} 
+ \mathbf{c}_{j}^\dag V_{ji}  \mathbf{c}_{i}, 
\Big]
\end{align}
where
$u_{ji}$ 
and $v_{ji}$ 
are complex numbers satisfying
\begin{align}
 u^*_{ji} = u_{ij}\,,v^*_{ji} = v_{ij}\,. \ \ \ \ 
\end{align}
To construct a $SU(2)$ SPT state using the above projective construction, we choose the ansatz $u_{ij}$ such that $f_\uparrow$ and $f_\downarrow$ fermions form two band insulators (each of which has $N_L$ bands, $N_L>1$). $f_\uparrow$  ($f_\downarrow$) fermions occupy one Chern band with Chern number $+1$ ($+1$). We also choose the ansatz $v_{ji}$ such that $c$ fermions (with two different "pseudo-spins") form a band insulator with $2N_L$ bands in total. The lowest two bands (each of them admits Chern number $-1$) are degenerate and filled by $c$ fermions with different pseudo-spins, respectively. Recently, some efforts have been made in lattice models with nontrivial Chern number larger than one, such as Refs.  \onlinecite{Wang11,YFWang12,Trescher12,Yang12}.   As a result, there are in total one $f_\uparrow$ fermion, one  $f_\downarrow$ fermion and two $c$\, fermions in each unit cell, required by the particle number constraint.

\subsection{Spinful SU(2) SPT state with charged superfluid}
In the
following, we like to show that $|\Phi\>=P|\Psi\>$ will be a bosonic SPT state
with the SU(2) symmetry. 
First, from the  free fermion Hamitonian, we see that the ground state state
$|\Psi\>$ respects the SU(2) spin rotation symmetry generated by $f^\dag \si^l
f$.  The $f$\, fermions form an ``integer quantum Hall'' state described by
$U^2(1)$ Chern-Simons theory
\begin{align}
 \cL&=
\frac{1}{4\pi} K_{IJ} a_{I\mu}\prt_\nu a_{J\la}\eps^{\mu\nu\la}
+\frac{1}{2\pi} q_{I} A^{\text{spin}}_{\mu}\prt_\nu a_{I\la}\eps^{\mu\nu\la}
\end{align}
with $I,J=1,2$ and
\begin{align}
 K=\bpm 1 &0 \\
        0 &1 
        \epm, \ \ \ \ 
 q=\bpm 1/2  \\
        -1/2 \epm
.
\end{align}
Here $A^{S_z}_\mu$ is the gauge potential that couples to the $S_z$ spin density
and current.  
The $c$\, fermions form an ``integer quantum Hall'' state described by
$U^2(1)$ Chern-Simons theory
\begin{align}
 \cL&=-
\frac{1}{4\pi} K_{IJ} b_{I\mu}\prt_\nu b_{J\la}\eps^{\mu\nu\la}
\end{align}
Thus the total effective theory is given by
\begin{align}
\cL&=
\frac{1}{4\pi} K_{IJ} a_{I\mu}\prt_\nu a_{J\la}\eps^{\mu\nu\la}
-\frac{1}{4\pi} K_{IJ} b_{I\mu}\prt_\nu b_{J\la}\eps^{\mu\nu\la}
\nonumber\\
&\ \ \ \ 
+\frac{1}{2\pi} q_{I} A^{\text{spin}}_{\mu}\prt_\nu a_{I\la}\eps^{\mu\nu\la}
\end{align}
The projection $P$ is done by setting the total $f$\, fermion
density-current, $J^f_\mu= \sum_I \frac{1}{2\pi} \prt_\nu a_{I\la} 
\eps^{\mu\nu\la} $, equal to the $c$\, fermion density-current, $J^c_\mu=
\sum_I \frac{1}{2\pi} \prt_\nu b_{I\la}  \eps^{\mu\nu\la} $.  After setting
$b_{2\la}=-b_{1\la}+\sum_I a_{I\la}$, we reduce the effective theory  to
\begin{align}
\label{SU2SPT1}
\cL&=
\frac{1}{4\pi} \bar K_{IJ} \bar a_{I\mu}\prt_\nu \bar a_{J\la}\eps^{\mu\nu\la}
+\frac{1}{2\pi} \bar q_{I} A^\text{spin}_{\mu}\prt_\nu \bar a_{I\la}\eps^{\mu\nu\la}
\end{align}
with $I,J=1,2,3$ and
\begin{align}
\bar K=
\bpm 
0 & -1 & 1  \\
-1 & 0 & 1  \\
1 & 1 & -2  \\
\epm, 
\ \ \ \
\bar q=\bpm 
1/2\\
-1/2\\
0
       \epm.
\end{align}
Also $\bar a_{I,\mu}=a_{I\mu}$ 
$I=1,2$,
and $\bar a_{3,\mu}=b_{1\mu}$.
Using an invertible integer matrix
\begin{align}
U=
\bpm 
1 & -1 & 0  \\
1 & 0 & 0  \\
1 & 1 & 1  \\
\epm
\end{align}
we can rewrite $\bar a_{I\mu}$ as $\bar a_{I\mu} =U^T \t a_{I\mu}$, and rewrite
\Eqn{SU2SPT1} as
\begin{align}
\label{SU2SPT}
\cL&=
\frac{1}{4\pi} \t K_{IJ} \t a_{I\mu}\prt_\nu \t a_{J\la}\eps^{\mu\nu\la}
+\frac{1}{2\pi} \t q_{I} A^\text{spin}_{\mu}\prt_\nu \t a_{I\la}\eps^{\mu\nu\la}
\end{align}
where
\begin{align}
\t K=U\bar KU^T=
\bpm 
2 & 1 &0 \\
1 & 0 &0 \\
0 & 0 &0 \\
\epm, 
\ \ \ \
\t q=U\bar q=\bpm 
1\\
1/2\\
0\\
       \epm.
\end{align}
\Eqn{SU2SPT} is the low energy effective theory for the spin-1/2 boson state
$|\Phi\>=P|\Psi\>$ after the projection.  We see that the mode $\t a_{3\mu}$ is
gapless. Such a gapless mode corresponds to the total density fluctuations of
the spin-1/2 bosons, indicating that the bosons are in a superfluid phase.  But
this is an unusual superfluid phase where the spin degrees of freedom form an
SU(2) SPT phase.

To see this point, let us assume that the
unit cell is large enough so that there are, on average, two spin-1/2 bosons
per unit cell.
In this case, when the repulsion between the bosons
is large enough, the bosons may form a Mott insulator state. 
Such a Mott insulator state is described by the
confinement of $\t a_{3\mu}$ U(1) gauge field. So we can drop
the $\t a_{3\mu}$ U(1) gauge field 
and obtain the following low energy effective theory
in the  Mott insulator phase:
\begin{align}
\label{SU2SPT2}
\cL&=
\frac{1}{4\pi} \t K_{2,IJ} \t a_{I\mu}\prt_\nu \t a_{J\la}\eps^{\mu\nu\la}
+\frac{1}{2\pi} \t q_{2,I} A^\text{spin}_{\mu}\prt_\nu \t a_{I\la}\eps^{\mu\nu\la}
\end{align}
where
\begin{align}
\t K_2=
\bpm 
2 & 1  \\
1 & 0  \\
\epm, 
\ \ \ \
\t q_2=\bpm 
1\\
1/2\\
       \epm.
\end{align}

Since det$(\t K_2)=-1$, the state $|\Phi\>$ (in the Mott insulator phase)
is a SPT state.
So, such a  low energy effective theory describes a
SU(2) SPT state, which has a spin Hall conductance
for the $S_z$ ``charge''
\begin{align}
 \si_{xy}=\t q_2^T\t K_2^{-1}\t q_2\frac{1}{2\pi}
 =2\frac{1}{4}\frac{1}{2\pi}.
\end{align}
We note that the  spin Hall conductance is an even integer
in the unit of $\frac{1}{4}\frac{1}{2\pi}$.
Although the above discussion is for spin-1/2 bosons, a similar
construction can be done for more physical spin-1 bosons, which
give us a SO(3) SPT state.

\subsection{Projective Kac-Moody algebra of the edge profile}\label{sectionKM}
Above quantum spin Hall effect gives the edge gapless Hall current protected by U(1) subgroup of SU(2). In the following, we shall construct the Kac-Moody algebra of edge profile which is actually described by SU(2)$_1$ WZW theory as same as Ref. (\onlinecite{Liu12}) where bulk non-linear sigma model is applied.
 
 Let's define the U(1)$_{L,R}$ scalar density operators $J_{L,R}$ and SU(2)$_{L,R}$ vector density operators $\mathbf{J}_{L,R}$ as 
\begin{align}
J_{L,R}=:\Psi^\dagger_{L,R;a}\Psi_{L,R;a}:\,\,,\\
\mathbf{J}_{L,R}=:\Psi^{\dagger}_{L,R;a}\frac{\vec{\sigma}_{ab}}{2}\Psi_{L,R;b}:\,\,, 
\end{align}
respectively, where, $\vec{\sigma}$ denotes a matrix vector formed by three Pauli matrices and $a,b,...=1,2$, and summation of common indices is implicit. $\Psi_{L}$ is a two-component gapless chiral fermion from the edge of $f$\, fermions; $\Psi_R$ is also a two-component gappless chiral fermion but from the edge of $c$\, fermions. ``$::$'' is usual normal ordering operator. The bosonized\cite{Witten84} edges admit U(1)$_{L,R}$$\times$SU(2)$_{L,R}$ Kac-Moody algebras as following commutation relations ($x,x'...$ are spatial coordinates of edge):
\begin{align}
[J_{L,R}(x),J_{L,R}(x')]=&\frac{1}{i\pi}\delta'(x-x')\,,\,\,\\
[J^{\alpha}_{L,R}(x),J^{\beta}_{L,R}(x')]=&i\epsilon^{\alpha\beta\gamma}J^{\gamma}_{L,R}(x)\delta(x-x')\nonumber\\
&-\frac{i}{4\pi}\delta^{\alpha\beta}\delta'(x-x')
\end{align}
The bosonized edge Hamiltonian consists of the energy of density-density interactions: 
\begin{align}
\mathcal{H}_{\text{edge}}=\sum_{s}\int dx  v_s\left[\frac{\pi}{2}:J_sJ_s:+\frac{2\pi}{3}:\mathbf{J}_s\cdot \mathbf{J}_s:\right]\,,\nonumber
\end{align}
where, $s=L,R$. $v_L$ and $v_R$ are the Fermi velocities of $\chi$FL (chiral Fermi liquid) of $f$ and $c$ systems, respectively, which are nonuniversal constants depending on microscopic details of projective Chern band parameters. The U(1) scalar density fluctuations must be gapped out by considering the Gutzwiller projection.\cite{Wen91PRB} The resultant edge theory at low-energies only consists the SU(2) vector density-density interaction terms with both of left and right movers:
\begin{align}
\mathcal{H}^{\text{Projected}}_{\text{edge}}=\sum_{s}\int dx  \frac{2\pi v_s}{3}:\mathbf{J}_s\cdot \mathbf{J}_s:\,\,\,,\nonumber
\end{align} 
Therefore, the Kac-Moody algebra is reconstructed after the projection as same as the result from non-linear sigma model in Ref. (\onlinecite{Liu12}).
 \section{Conclusion}
In conclusion, a general approach is proposed to construct bosonic SPT
phases on 2D lattice with various symmetries.  Our approach is based on the
projective construction where the bosons/spins are fractionalized into several
fermions which occupy nontrivial Chern bands.  The  bosonic SPT phases then
can be constructed from the fermion state by Gutzwiller projection.  We can
calculate the low energy effective Chern-Simons theory of the projected states,
which allows us to determine what kinds of SPT states are obtained after the
Gutzwiller projection.  We have constructed a U(1) SPT state, a SO(3) SPT state,
and a SU(2) SPT state for spin-1 and spin-1/2 bosons. In particular, in the SU(2) SPT phases, the U(1) charge is in gapless superfluid phase while the spin degree of freedom is in SU(2) SPT phase. 

The present approach can be generalized to more complex continuous nonabelian symmetry group, such as SU(N) SPT magnets. Moreover, generally the same approach can be applied to construct SET phases which have intrinsic topological order (with $|\text{det}K|\neq1$) protected by symmetry. Finally, the present general approach provides a systematical way to construct the trial projective wavefunctions of 2D SPT states. In the future work, it is quite interesting to search concrete microscopic spin/boson Hamiltonians that realize the mean-field ansatzes and trial wavefunctions. And, many projective wavefunctions can be constructed in the approach such that it will be attractive to numerically study the ground state properties via numerical methods such as variational Monte Carlo.

\section*{Acknowledgement} We would like to thank Cenke Xu, Meng Cheng, Shinsei Ryu, Yi Zhang and Xiao-Liang Qi's helpful discussions.  This
research is supported by NSF Grant No. DMR-1005541, NSFC 11074140, and NSFC
11274192. (XGW) Research at Perimeter Institute is supported by the Government of
Canada through Industry Canada and by the Province of Ontario through the
Ministry of Research and Innovation.  (PY and XGW)

%\bibliography{../../bib/wencross,../../bib/all,../../bib/publst,./tmp}

\begin{thebibliography}{26}
\expandafter\ifx\csname natexlab\endcsname\relax\def\natexlab#1{#1}\fi
\expandafter\ifx\csname bibnamefont\endcsname\relax
  \def\bibnamefont#1{#1}\fi
\expandafter\ifx\csname bibfnamefont\endcsname\relax
  \def\bibfnamefont#1{#1}\fi
\expandafter\ifx\csname citenamefont\endcsname\relax
  \def\citenamefont#1{#1}\fi
\expandafter\ifx\csname url\endcsname\relax
  \def\url#1{\texttt{#1}}\fi
\expandafter\ifx\csname urlprefix\endcsname\relax\def\urlprefix{URL }\fi
\providecommand{\bibinfo}[2]{#2}
\providecommand{\eprint}[2][]{\url{#2}}

\bibitem[{\citenamefont{Levin and Wen}(2006)}]{LW0605}
\bibinfo{author}{\bibfnamefont{M.}~\bibnamefont{Levin}} \bibnamefont{and}
  \bibinfo{author}{\bibfnamefont{X.-G.} \bibnamefont{Wen}},
  \bibinfo{journal}{Phys. Rev. Lett.} \textbf{\bibinfo{volume}{96}},
  \bibinfo{pages}{110405} (\bibinfo{year}{2006}), \eprint{cond-mat/0510613}.

\bibitem[{\citenamefont{Kitaev and Preskill}(2006)}]{KP0604}
\bibinfo{author}{\bibfnamefont{A.}~\bibnamefont{Kitaev}} \bibnamefont{and}
  \bibinfo{author}{\bibfnamefont{J.}~\bibnamefont{Preskill}},
  \bibinfo{journal}{Phys. Rev. Lett.} \textbf{\bibinfo{volume}{96}},
  \bibinfo{pages}{110404} (\bibinfo{year}{2006}).


\bibitem{Chenlong} Xie Chen, Z. C. Gu, Z. X. Liu, and X. G. Wen, arxiv:1106.4772.
\bibitem{Chen10} Xie Chen, Z. C. Gu, and X. G. Wen, Phys. Rev. B \textbf{82}, 155138 (2010).

  \bibitem{Wenscience} X. Chen, Z.-C. Gu, Z.-X. Liu, X.-G. Wen, Science 338, 1604 (2012); X. L. Qi, Science 338, 1550 (2012).

\bibitem{Tsui82} D. C. Tsui, H. L. Stormer, and A. C. Gossard, Phys. Rev. Lett. \textbf{48}, 1559 (1982).

\bibitem[{\citenamefont{Wen}(1989)}]{Wtop}
\bibinfo{author}{\bibfnamefont{X.-G.} \bibnamefont{Wen}},
  \bibinfo{journal}{Phys. Rev. B} \textbf{\bibinfo{volume}{40}},
  \bibinfo{pages}{7387} (\bibinfo{year}{1989}).

\bibitem[{\citenamefont{Wen and Niu}(1990)}]{WNtop}
\bibinfo{author}{\bibfnamefont{X.-G.} \bibnamefont{Wen}} \bibnamefont{and}
  \bibinfo{author}{\bibfnamefont{Q.}~\bibnamefont{Niu}},
  \bibinfo{journal}{Phys. Rev. B} \textbf{\bibinfo{volume}{41}},
  \bibinfo{pages}{9377} (\bibinfo{year}{1990}).

\bibitem[{\citenamefont{Wen}(1990)}]{Wrig}
\bibinfo{author}{\bibfnamefont{X.-G.} \bibnamefont{Wen}},
  \bibinfo{journal}{Int. J. Mod. Phys. B} \textbf{\bibinfo{volume}{4}},
  \bibinfo{pages}{239} (\bibinfo{year}{1990}).
  \bibitem{TI} M. Z. Hasan, C. L. Kane, Rev. Mod. Phys. 82, 3045
(2010); J. E. Moore, Nature 464, 194 (2010); X.-L. Qi, S.-C. Zhang, Rev. Mod. Phys. 83, 1057 (2011).
%\bibitem{TI} C. L. Kane and E. J. Mele, Phys. Rev. Lett. \textbf{95}, 146802 
%(2005); B. A. Bernevig and S.-C. Zhang, Phys. Rev. Lett. \textbf{96}, 106802 (2006);  J. E. Moore and L. Balents, Phys. Rev. B \textbf{75}, 121306 (2007);  L. Fu, C. L. Kane, and E. J. Mele, Phys. Rev. Lett. \textbf{98}, 106803 (2007); X.-L. Qi, T. Hughes, and S.-C. Zhang, Phys. Rev. B \textbf{78},195424 (2008); X. L. Qi and S. C. Zhang, Rev. Mod. Phys. \textbf{83}, 1057 (2011).
\bibitem{Chen11a} Xie Chen, Z. X. Liu, and X. G. Wen, Phys. Rev. B \textbf{84}, 235141 (2011).
\bibitem{Liu11}Z. X. Liu, X. Chen, and X. G. Wen, Phys. Rev. B \textbf{84}, 195145 (2011).


\bibitem[{\citenamefont{Lu and Vishwanath}(2012)}]{LV1219}
\bibinfo{author}{\bibfnamefont{Y.-M.} \bibnamefont{Lu}} \bibnamefont{and}
  \bibinfo{author}{\bibfnamefont{A.}~\bibnamefont{Vishwanath}},
  \bibinfo{journal}{Phys. Rev.} \textbf{\bibinfo{volume}{86}},
  \bibinfo{pages}{125119} (\bibinfo{year}{2012}), \eprint{arXiv:1205.3156}.

\bibitem[{\citenamefont{Levin and Stern}(2009)}]{LS0903}
\bibinfo{author}{\bibfnamefont{M.}~\bibnamefont{Levin}} \bibnamefont{and}
  \bibinfo{author}{\bibfnamefont{A.}~\bibnamefont{Stern}},
  \bibinfo{journal}{Phys. Rev. Lett.} \textbf{\bibinfo{volume}{103}},
  \bibinfo{pages}{196803} (\bibinfo{year}{2009}), \eprint{arXiv:0906.2769}.

\bibitem{Liu12} Zheng-Xin Liu and Xiao-Gang Wen, arxiv:1205.7024.
\bibitem{WZ71} J. Wess and B. Zumino, Phys. Lett. B \textbf{37}, 95 (1971).
\bibitem{Witten83} E. Witten, Nucl. Phys. B \textbf{223}, 422 (1983).
\bibitem{Witten84}E. Witten, Commun. Math. Phys. \textbf{92}, 455 (1984).


 
\bibitem[{\citenamefont{Baskaran et~al.}(1987)\citenamefont{Baskaran, Zou, and
  Anderson}}]{BZA8773}
\bibinfo{author}{\bibfnamefont{G.}~\bibnamefont{Baskaran}},
  \bibinfo{author}{\bibfnamefont{Z.}~\bibnamefont{Zou}}, \bibnamefont{and}
  \bibinfo{author}{\bibfnamefont{P.~W.} \bibnamefont{Anderson}},
  \bibinfo{journal}{Solid State Comm.} \textbf{\bibinfo{volume}{63}},
  \bibinfo{pages}{973} (\bibinfo{year}{1987}).

\bibitem[{\citenamefont{Baskaran and Anderson}(1988)}]{BA8880}
\bibinfo{author}{\bibfnamefont{G.}~\bibnamefont{Baskaran}} \bibnamefont{and}
  \bibinfo{author}{\bibfnamefont{P.~W.} \bibnamefont{Anderson}},
  \bibinfo{journal}{Phys. Rev. B} \textbf{\bibinfo{volume}{37}},
  \bibinfo{pages}{580} (\bibinfo{year}{1988}).

\bibitem[{\citenamefont{Affleck and Marston}(1988)}]{AM8874}
\bibinfo{author}{\bibfnamefont{I.}~\bibnamefont{Affleck}} \bibnamefont{and}
  \bibinfo{author}{\bibfnamefont{J.~B.} \bibnamefont{Marston}},
  \bibinfo{journal}{Phys. Rev. B} \textbf{\bibinfo{volume}{37}},
  \bibinfo{pages}{3774} (\bibinfo{year}{1988}).

\bibitem[{\citenamefont{Kotliar and Liu}(1988)}]{KL8842}
\bibinfo{author}{\bibfnamefont{G.}~\bibnamefont{Kotliar}} \bibnamefont{and}
  \bibinfo{author}{\bibfnamefont{J.}~\bibnamefont{Liu}},
  \bibinfo{journal}{Phys. Rev. B} \textbf{\bibinfo{volume}{38}},
  \bibinfo{pages}{5142} (\bibinfo{year}{1988}).

\bibitem[{\citenamefont{Suzumura et~al.}(1988)\citenamefont{Suzumura, Hasegawa,
  and Fukuyama}}]{SHF8868}
\bibinfo{author}{\bibfnamefont{Y.}~\bibnamefont{Suzumura}},
  \bibinfo{author}{\bibfnamefont{Y.}~\bibnamefont{Hasegawa}}, \bibnamefont{and}
  \bibinfo{author}{\bibfnamefont{H.}~\bibnamefont{Fukuyama}},
  \bibinfo{journal}{J. Phys. Soc. Jpn.} \textbf{\bibinfo{volume}{57}},
  \bibinfo{pages}{2768} (\bibinfo{year}{1988}).

\bibitem[{\citenamefont{Affleck et~al.}(1988)\citenamefont{Affleck, Zou, Hsu,
  and Anderson}}]{AZH8845}
\bibinfo{author}{\bibfnamefont{I.}~\bibnamefont{Affleck}},
  \bibinfo{author}{\bibfnamefont{Z.}~\bibnamefont{Zou}},
  \bibinfo{author}{\bibfnamefont{T.}~\bibnamefont{Hsu}}, \bibnamefont{and}
  \bibinfo{author}{\bibfnamefont{P.~W.} \bibnamefont{Anderson}},
  \bibinfo{journal}{Phys. Rev. B} \textbf{\bibinfo{volume}{38}},
  \bibinfo{pages}{745} (\bibinfo{year}{1988}).

\bibitem[{\citenamefont{Dagotto et~al.}(1988)\citenamefont{Dagotto, Fradkin,
  and Moreo}}]{DFM8826}
\bibinfo{author}{\bibfnamefont{E.}~\bibnamefont{Dagotto}},
  \bibinfo{author}{\bibfnamefont{E.}~\bibnamefont{Fradkin}}, \bibnamefont{and}
  \bibinfo{author}{\bibfnamefont{A.}~\bibnamefont{Moreo}},
  \bibinfo{journal}{Phys. Rev. B} \textbf{\bibinfo{volume}{38}},
  \bibinfo{pages}{2926} (\bibinfo{year}{1988}).

\bibitem[{\citenamefont{Wen et~al.}(1989)\citenamefont{Wen, Wilczek, and
  Zee}}]{WWZcsp}
\bibinfo{author}{\bibfnamefont{X.-G.} \bibnamefont{Wen}},
  \bibinfo{author}{\bibfnamefont{F.}~\bibnamefont{Wilczek}}, \bibnamefont{and}
  \bibinfo{author}{\bibfnamefont{A.}~\bibnamefont{Zee}},
  \bibinfo{journal}{Phys. Rev. B} \textbf{\bibinfo{volume}{39}},
  \bibinfo{pages}{11413} (\bibinfo{year}{1989}).

\bibitem[{\citenamefont{Wen}(1991)}]{Wsrvb}
\bibinfo{author}{\bibfnamefont{X.-G.} \bibnamefont{Wen}},
  \bibinfo{journal}{Phys. Rev. B} \textbf{\bibinfo{volume}{44}},
  \bibinfo{pages}{2664} (\bibinfo{year}{1991}).

\bibitem[{\citenamefont{Lee and Nagaosa}(1992)}]{LN9221}
\bibinfo{author}{\bibfnamefont{P.~A.} \bibnamefont{Lee}} \bibnamefont{and}
  \bibinfo{author}{\bibfnamefont{N.}~\bibnamefont{Nagaosa}},
  \bibinfo{journal}{Phys. Rev. B} \textbf{\bibinfo{volume}{45}},
  \bibinfo{pages}{5621} (\bibinfo{year}{1992}).

\bibitem[{\citenamefont{Mudry and Fradkin}(1994)}]{MF9400}
\bibinfo{author}{\bibfnamefont{C.}~\bibnamefont{Mudry}} \bibnamefont{and}
  \bibinfo{author}{\bibfnamefont{E.}~\bibnamefont{Fradkin}},
  \bibinfo{journal}{Phys. Rev. B} \textbf{\bibinfo{volume}{49}},
  \bibinfo{pages}{5200} (\bibinfo{year}{1994}).

\bibitem[{\citenamefont{Wen and Lee}(1996)}]{WLsu2}
\bibinfo{author}{\bibfnamefont{X.-G.} \bibnamefont{Wen}} \bibnamefont{and}
  \bibinfo{author}{\bibfnamefont{P.~A.} \bibnamefont{Lee}},
  \bibinfo{journal}{Phys. Rev. Lett.} \textbf{\bibinfo{volume}{76}},
  \bibinfo{pages}{503} (\bibinfo{year}{1996}), \eprint{cond-mat/9506065}.
  
 \bibitem{YW13} Peng Ye and X.-G. Wen, arxiv:1303.3572. 
  

\bibitem[{\citenamefont{Chen and Wen}(2012)}]{CW1217}
\bibinfo{author}{\bibfnamefont{X.}~\bibnamefont{Chen}} \bibnamefont{and}
  \bibinfo{author}{\bibfnamefont{X.-G.} \bibnamefont{Wen}}
  (\bibinfo{year}{2012}), \eprint{arXiv:1206.3117}.

\bibitem[{\citenamefont{Senthil and Levin}(2012)}]{SL1204}
\bibinfo{author}{\bibfnamefont{T.}~\bibnamefont{Senthil}} \bibnamefont{and}
  \bibinfo{author}{\bibfnamefont{M.}~\bibnamefont{Levin}}
  (\bibinfo{year}{2012}), \eprint{arXiv:1206.1604}.

\bibitem[{\citenamefont{Grover and Vishwanath}(2012)}]{GV1207}
\bibinfo{author}{\bibfnamefont{T.}~\bibnamefont{Grover}} \bibnamefont{and}
  \bibinfo{author}{\bibfnamefont{A.}~\bibnamefont{Vishwanath}}
  (\bibinfo{year}{2012}), \eprint{arXiv:1210.0907}.
\bibitem{Lu12b} Y.-M. Lu and D.-H. Lee (2012), arXiv:1210.0909.
\bibitem[{\citenamefont{Lu and Lee}(2012)}]{LL1263}
\bibinfo{author}{\bibfnamefont{Y.-M.} \bibnamefont{Lu}} \bibnamefont{and}
  \bibinfo{author}{\bibfnamefont{D.-H.} \bibnamefont{Lee}}
  (\bibinfo{year}{2012}), \eprint{arXiv:1212.0863}.
   
\bibitem[{\citenamefont{Kane and Mele}(2005)}]{KM0501}
\bibinfo{author}{\bibfnamefont{C.~L.} \bibnamefont{Kane}} \bibnamefont{and}
  \bibinfo{author}{\bibfnamefont{E.~J.} \bibnamefont{Mele}},
  \bibinfo{journal}{Phys. Rev. Lett.} \textbf{\bibinfo{volume}{95}},
  \bibinfo{pages}{226801} (\bibinfo{year}{2005}), \eprint{cond-mat/0411737}.

\bibitem[{\citenamefont{Wen}(1995)}]{Wtoprev}
\bibinfo{author}{\bibfnamefont{X.-G.} \bibnamefont{Wen}},
  \bibinfo{journal}{Advances in Physics} \textbf{\bibinfo{volume}{44}},
  \bibinfo{pages}{405} (\bibinfo{year}{1995}).

\bibitem[{\citenamefont{Wen}(1999)}]{Wpcon}
\bibinfo{author}{\bibfnamefont{X.-G.} \bibnamefont{Wen}},
  \bibinfo{journal}{Phys. Rev. B} \textbf{\bibinfo{volume}{60}},
  \bibinfo{pages}{8827} (\bibinfo{year}{1999}), \eprint{cond-mat/9811111}.

%\bibitem{Auerbach} Assa Auerbach, \textit{Interacting Electrons and Quantum Magnetism} (Springer-Verlag, 1994).
%
%\bibitem{Tang11}Evelyn Tang, Jia-Wei Mei, and Xiao-Gang Wen, Phys. Rev. Lett. \textbf{106}, 236802 (2011).
%\bibitem{Sun11} Kai Sun, Zhengcheng Gu, Hosho Katsura, and S. Das Sarma, Phys. Rev. Lett. \textbf{106}, 236803 (2011).
%\bibitem{Neupert11} Titus Neupert, Luiz Santos, Claudio Chamon, and Christopher Mudry, Phys. Rev. Lett. \textbf{106}, 236804 (2011).
%\bibitem{Grushin12} Adolfo G. Grushin, Titus Neupert, Claudio Chamon, Christopher Mudry, Phys. Rev. B \textbf{86} 205125 (2012).
\bibitem{Yang12} Shuo Yang, Zheng-Cheng Gu, Kai Sun, and S. Das Sarma, arxiv:1205.5792.
\bibitem{Wang11} F. Wang and Y. Ran, Phys. Rev. B \textbf{84}, 241103 (2011).
\bibitem{YFWang12} Yi-Fei Wang, Hong Yao, Chang-De Gong, and D. N. Sheng, Phys. Rev. B 86, 201101 (2012).
\bibitem{Trescher12}Maximilian Trescher and Emil J. Bergholtz, Phys. Rev. B 86, 241111 (2012).
\bibitem{Wen91PRB} X.-G. Wen, Phys. Rev. B \textbf{43} 11025, (1991).
%\bibitem{Lu12} Y. M. Lu and D. H. Lee, arxiv:1210.0909.
%\bibitem{Grover12}T. Grover and A. Vishwanath, arxiv:1210.0907.
%\bibitem{Wen91} X.-G. Wen, Mod. Phys. Lett. B \textbf{5}, 39 (1991).
%

%\bibitem{Halperin82} B. I. Halperin, Phys. Rev. B \textbf{25}, 2185 (1982).
%
%\bibitem{Wen90} X.-G. Wen, Phys. Rev. B \textbf{41}, 12838 (1990).
%
%\bibitem{Kh89}D. V. Khveshchenko and P. Wiegmann, Mod. Phys. Lett. B \textbf{3}, 1383 (1989).
%\bibitem{Wen89}X.-G. Wen, F. Wilczek, and A. Zee, Phys. Rev. B \textbf{39}, 11413 (1989).
%\bibitem{Affleck89} I. Affleck, J. Harvey, L. Palla, and G. Semenoff, Nucl. Phys. B \textbf{328}, 575 (1989).
%\bibitem{Fradkin91} E. Fradkin and F. A. Schaposnik, Phys. Rev. Lett. \textbf{66}, 276 (1991).
%\bibitem{Kogan96} I. I. Kogan and A. Kovner, Phys. Rev. D \textbf{53}, 4510 (1996).
%\bibitem{Smorgrav04}E. Smorgrav, J. Smiseth, A. Sudbo and F. S. Nogueira, Europhys. Lett. \textbf{68}, 198 (2004).
%\bibitem{Polyakov75} A. M. Polyakov, Phys. Lett. B \textbf{59}, 82 (1975).
%\bibitem{Polyakovpaper}A. M. Polyakov, Nucl. Phys. B \textbf{120}, 429 (1977).
%\bibitem{Polyakovbook} A. M. Polyakov, \emph{ Gauge Fields and Strings} (Harwood Academic, New York, 1987).
%\bibitem{Deser82} S. Deser, R. Jackiw, and S. Templeton, Ann. Phys. \textbf{140}, 372 (1982).
% 
%\bibitem{Levin12} M. Levin and Z. C. Gu, arxiv:1202.3120.
%\bibitem{Affleck87} I. Affleck and F. D. M. Haldane, Phys. Rev. B \textbf{36}, 5291 (1987).
%\bibitem{Nielsen11} A. E. B. Nielsen, J. I. Cirac, and G. Sierra, J. Stat. Mech. (2011)
%P11014.
%\bibitem{Thomale11} R. Thomale, S. Rachel, P. Schmitteckert, and M. Greiter, Phys. Rev. B \textbf{85}, 195149 (2012).
%\bibitem{Tu12} Hong-Hao Tu, arxiv:1210.1481.
%\bibitem{AKLT} I. Affleck, T. Kennedy, E. H. Lieb, and H. Tasaki, Phys. Rev. Lett. \textbf{59}, 799 (1987).






\end{thebibliography}

\end{document}